\documentclass[aps,prb,twocolumn,float,amsmath]{revtex4}
\usepackage[dvips]{color}
\usepackage[normalem]{ulem} 
 \usepackage{hyperref}
\definecolor{g-blue}{rgb}{0.83,0.95,1}
\definecolor{g-yellow}{rgb}{1,1,0.7}
\definecolor{g-green}{rgb}{0.9,1,0.9}
\definecolor{green}{rgb}{0,0.6,0}
\definecolor{cyan}{rgb}{0,0.7,0.7}
\definecolor{black}{rgb}{0,0,0}
\definecolor{grey}{rgb}{0.4 ,0.4 ,0.4 }

\usepackage{bm}
\usepackage{amssymb}
\usepackage{amsfonts}
\usepackage{amsmath}
 \usepackage{graphicx}
  \usepackage{amsmath,bm,epsfig}
\def \ed {\end{document}}
\def\Fbox#1{\vskip1ex\hbox to 8.5cm{\hfil\fboxsep0.3cm\fbox{%
  \parbox{8.0cm}{#1}}\hfil}\vskip1ex\noindent}  %%  {TEXT} in BOX

\newcommand{\eq}[1]{(\ref{#1})}%%  requires \eq{label}
\newcommand{\Eq}[1]{Eq.\,(\ref{#1})}%%  requires \eq{label}
\newcommand{\Eqs}[1]{Eqs.\,(\ref{#1})}%%  requires \eq{label}
\newcommand{\Fig}[1]{Fig.\,\ref{#1}}%%  requires \Fef{label}
\newcommand{\Sec}[1]{Sec.\,\ref{#1}}%%  requires \Fef{label}
\newcommand{\Ref}[1]{Ref.\,\cite{#1}}%%  requires \Fef{label}
\newcommand{\Refs}[1]{Refs.\,\cite{#1}}%%  requires \Fef{label}

\def\be{\begin{equation}}\def\ee{\end{equation}}
\def\bea{\begin{eqnarray}}\def\eea{\end{eqnarray}}
\def\bse{\begin{subequations}}\def\ese{\end{subequations}}
\newcommand{\BE}[1]{\begin{equation}\label{#1}}
\newcommand{\BEA}[1]{\begin{eqnarray}\label{#1}}
\newcommand{\BSE}[1]{\begin{subequations}\label{#1}}

\let \nn  \nonumber  \newcommand{\br}{\\ \nn}

\let \= \equiv \let\*\cdot \let\~\widetilde \let\^\widehat \let\-\overline
\let\p\partial

  \def\1{\bm1} 

\def\<{\left\langle}    \def\>{\right\rangle}
\def\({\left(}          \def\){\right)}
 \def \[ {\left [} \def \] {\right ]}

\renewcommand{\a}{\alpha}\newcommand{\g}{\gamma}
\newcommand{\G} {\Gamma}\renewcommand{\d}{\delta}
\newcommand{\D}{\Delta}
\renewcommand{\o}{\omega} \renewcommand{\O}{\Omega}

\newcommand{\B}[1]{{\bm{#1}}}%% Bold Roman & Greek Lower & Upper Case
\newcommand{\C}[1]{{\mathcal{#1}}}    %%   Calligrapfic Upper case
\renewcommand{\sb}[1]{_{\text {#1}}}  %% sub-   for lower case
\renewcommand{\sp}[1]{^{\text {#1}}}  %% super- for lower case
\newcommand{\Sp}[1]{^{^{\text {#1}}}} %% Super- for Upper case

\def\He4 {$^4$He~}

\begin{document}

\title{Counter-flow Induced Decoupling in Super-Fluid Turbulence}
\author{Dmytro Khomenko, Victor S. L'vov,   Anna Pomyalov, and Itamar
Procaccia }
\affiliation{Department of Chemical Physics,  Weizmann Institute  of Science, Rehovot 76100, Israel }

\begin{abstract} In mechanically driven  superfluid turbulence the mean velocities of the normal- and superfluid components  are known to coincide: $\B U\sb n =\B U\sb s$. Numerous laboratory, numerical and analytical studies showed that under these conditions the mutual friction between the normal- and superfluid velocity components couples also their fluctuations:  $\B u'\sb n(\B r,t) \approx \B u'\sb s(\B r,t)$ almost at all scales. In this paper we show that this is not the case in thermally driven superfluid turbulence; here the counterflow velocity $\B U\sb {ns}\equiv \B U\sb n -\B U\sb s\ne 0$. We suggest a simple analytic model for the cross correlation function $\< \B u'\sb n(\B r,t) \cdot  \B u'\sb s(\B r',t)\>$ and its dependence on $U\sb{ns}$. We demonstrate that  $\B u'\sb n(\B r,t)$ and $ \B u'\sb s(\B r,t)$ are  decoupled almost in the entire range of separations $|\B r-\B r'|$ between the energy containing scale and intervortex distance.
\end{abstract}

\maketitle

\section{\label{s:intro}Introduction}
%--------------------------------------------

\begin{figure*}%[h!]
 \begin{tabular}{c c c}
& Co-flow  & \\
  (a)  $t=-\tau$ & (b) $t=0$ & (c) $t=\tau$ \\
   \includegraphics[width=5.6 cm, keepaspectratio]{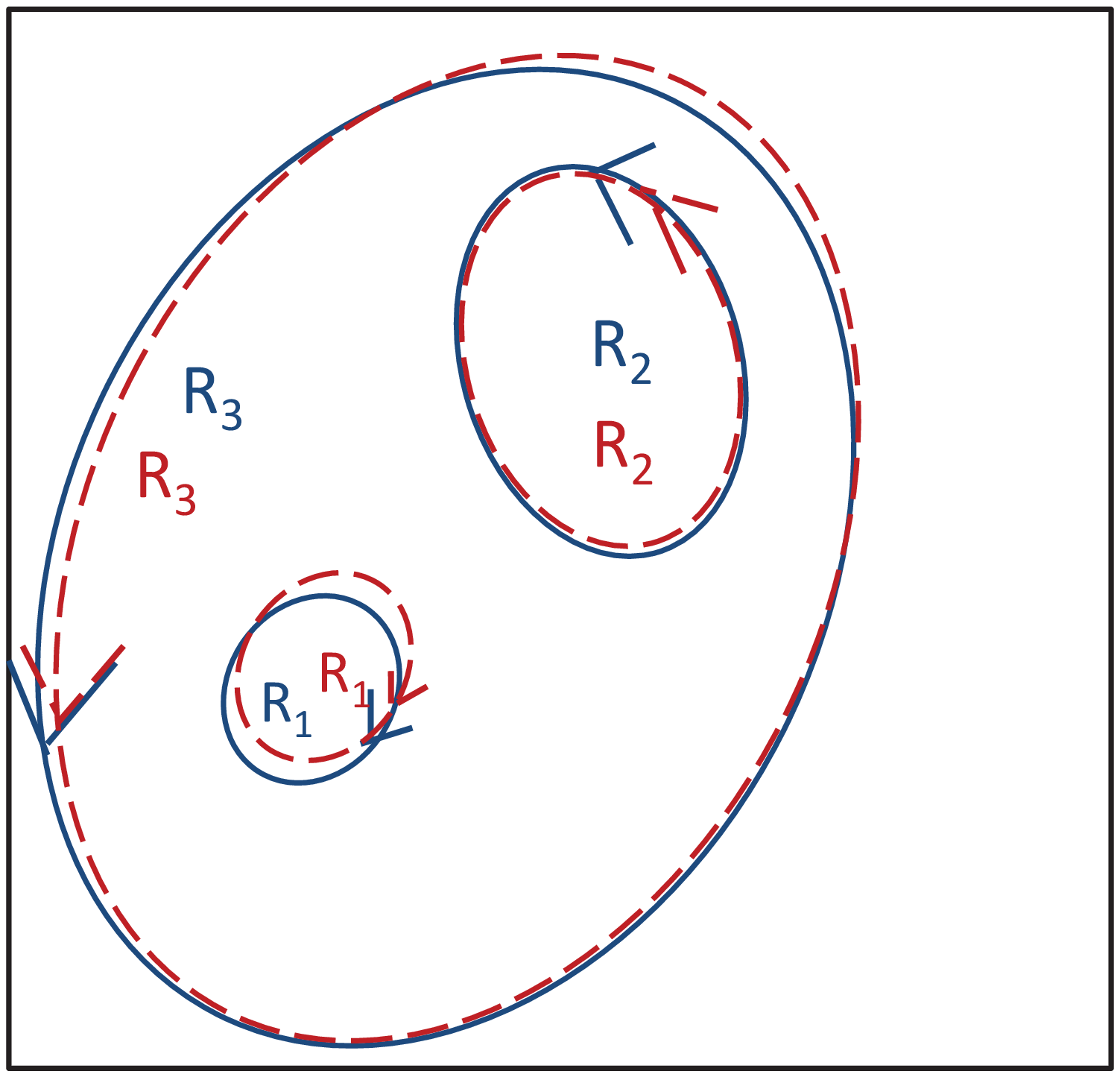} &
 \includegraphics[width=5.6 cm, keepaspectratio]{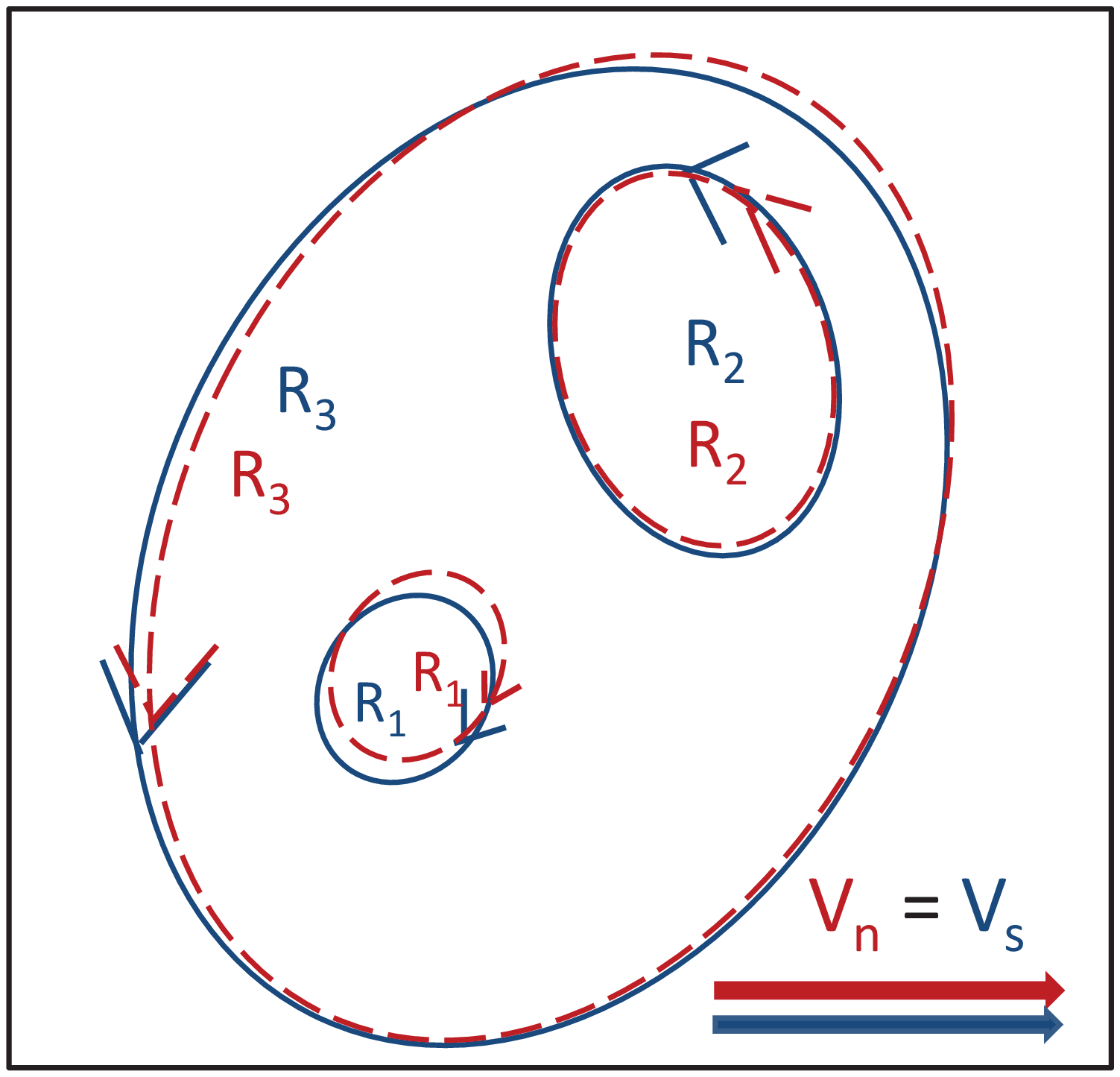} &
 \includegraphics[width=5.6 cm, keepaspectratio]{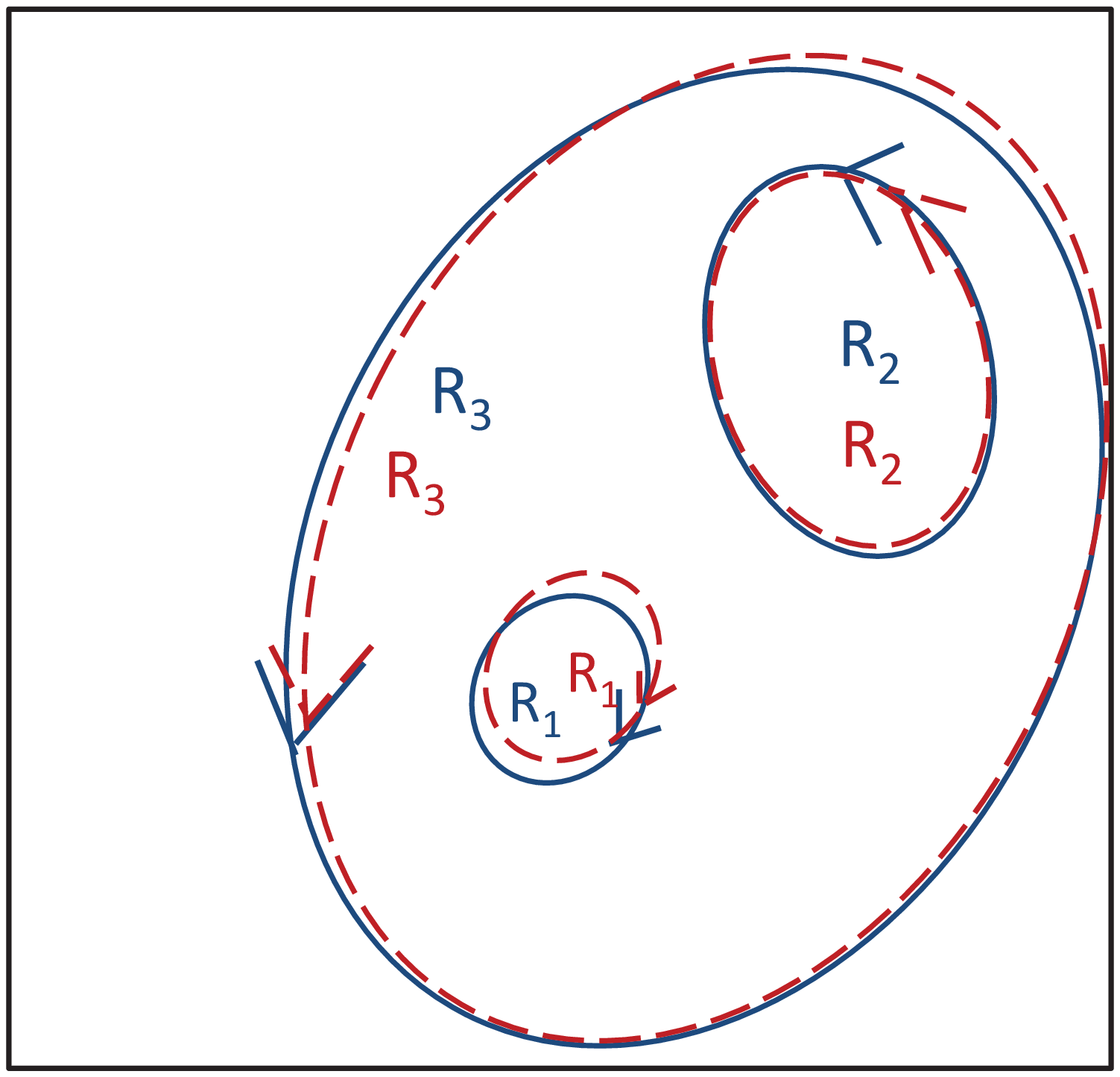}\\
 & Counter-flow  & \\
  (d)  $t=-\tau$  & (e) $t=0$& (f) $t=\tau$\\
  \includegraphics[width=5.6 cm, keepaspectratio]{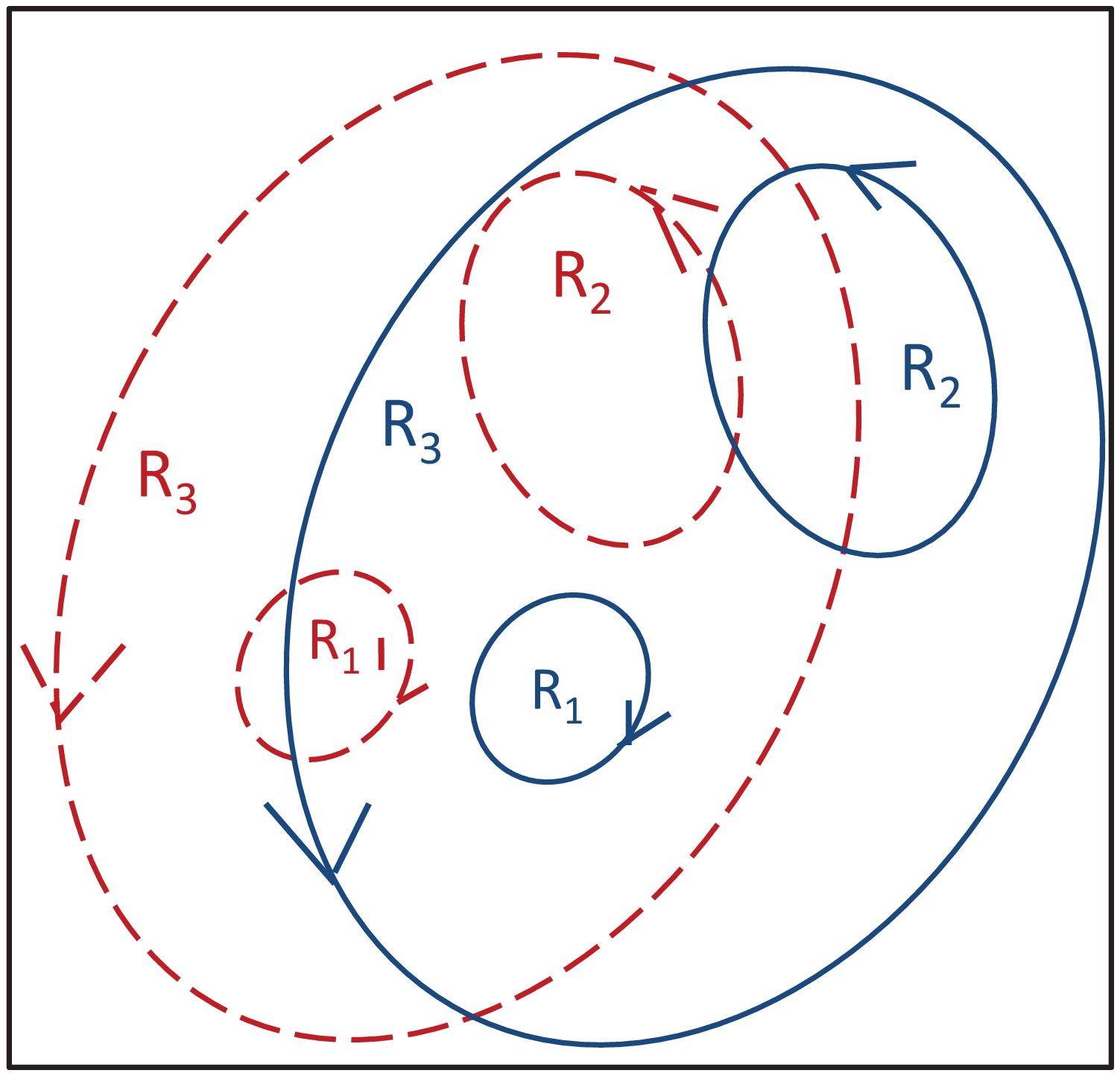}&
 \includegraphics[width=5.6 cm, keepaspectratio]{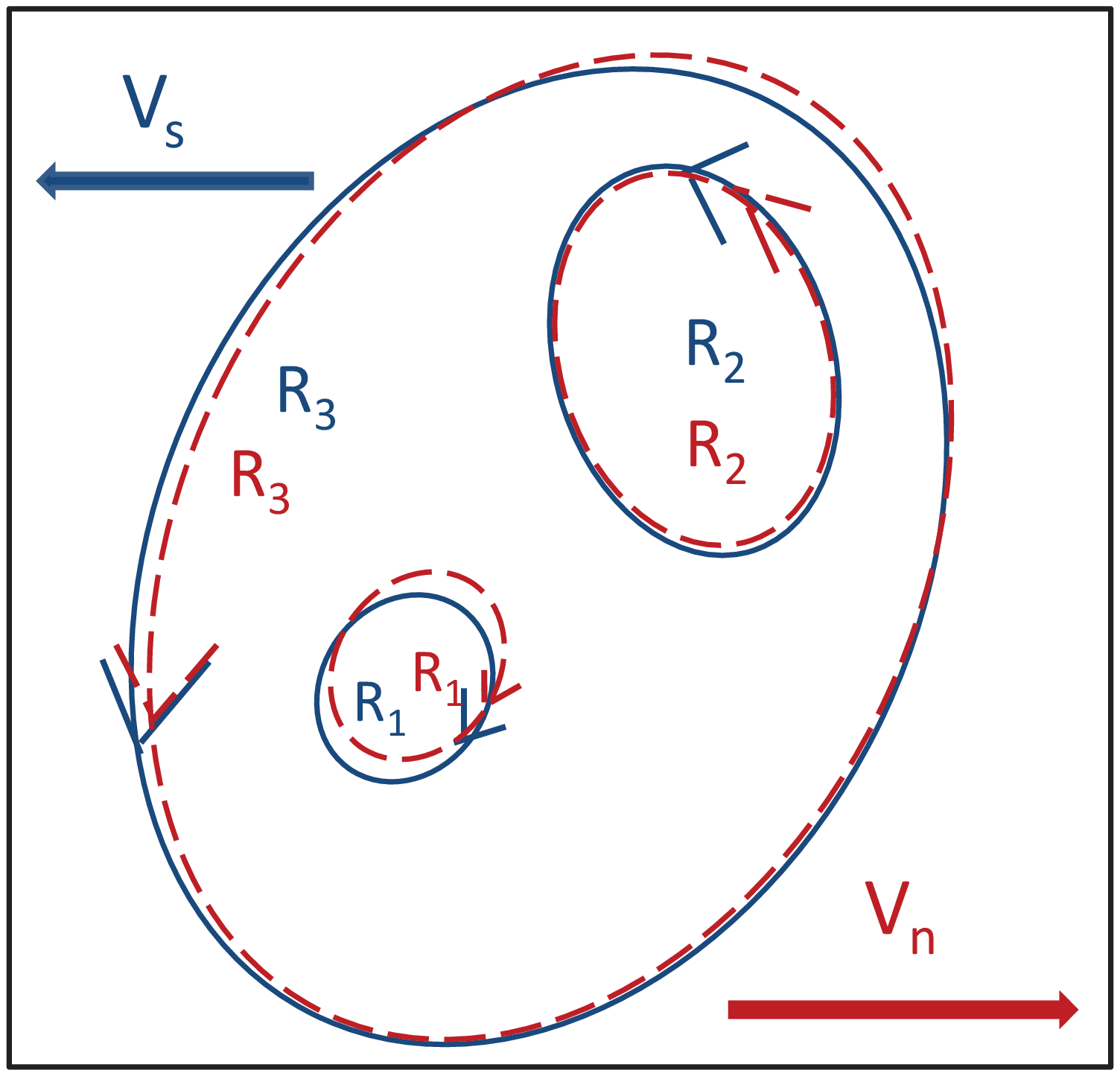} &
 \includegraphics[width=5.6 cm, keepaspectratio]{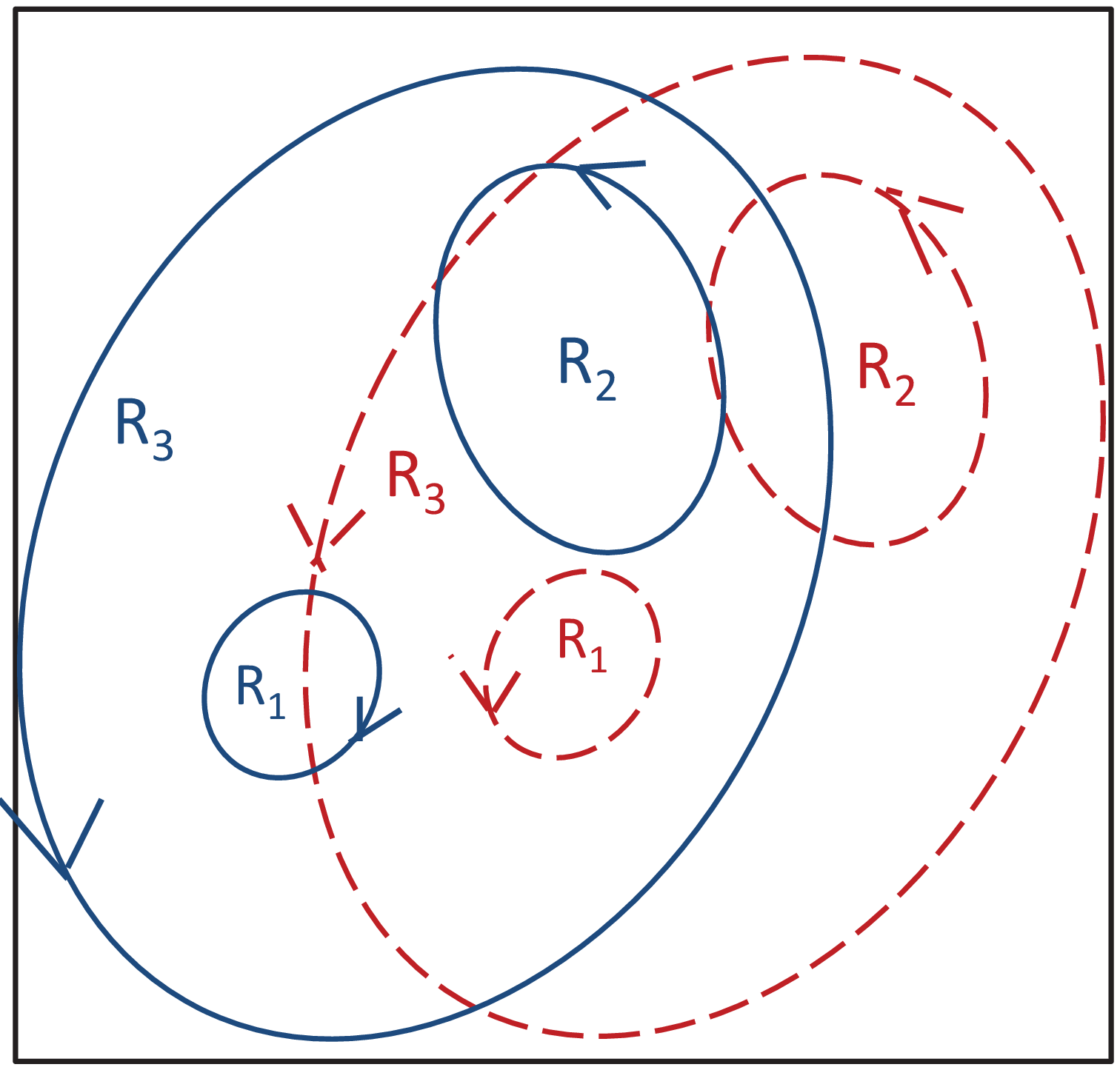}\\
 \end{tabular}
 \caption{\label{F:1} Color online.  Schematic view of the normal fluid eddies of scales
  $R_1$, $R_2$ and $R_3$ (shown by red dashed lines), swept by the mean
  normal fluid velocity $U\sb n$, and of the superfluid eddies of the same scales
   (shown by blue solid lines) swept by the mean superfluid
   velocity $U\sb s$ in the co-flow with $\B U\sb n= \B U\sb s$ [panels (a), (b) and (c)]
   and in  the counter-flow with $U\sb {ns}=|\B U\sb n - \B U\sb s|\ne 0$ [panels (d), (e) and (f)] at three
   consequent moment of times: $t=-\tau$ [panels (a) and (d)], $t=0$
    [panels (b) and (e)] and $t=\tau$  [panels (c) and (f)]. The time interval $\tau\simeq R_2/ U\sb {ns}$ is
     of the order of overlapping time of the middle-scale  $R_2$-eddies. }
\end{figure*}
%--------------------------------------------
Much of the thinking about turbulence in quantum fluids like $^4$He at low temperature is still influenced by
the  ``two fluid" model of Landau and Tisza. Within this model
the dynamics of the superfluid $^4$He is described in terms of a viscous normal component and an inviscid superfluid component, each with its own density $\rho\sb n(T)$ and $\rho\sb s(T)$ and its own velocity field $\B u\sb n(\B r,t)$ and $\B u\sb s(\B r,t)$. Due to the quantum mechanical restriction, the circulation around the superfluid vortices is quantized to integer values of $\kappa= h /m$, where $h$ is the Plank constant and $m$ is the mass of $^4$He atom. The quantization of circulation results in the appearance of characteristic ``quantum" length scale: the mean separation between vortex lines, $\ell$, which is typically orders of magnitude smaller than the scale $H$ of the largest (energy containing)
eddies\,\cite{1,2}.

Experimental evidence \cite{SS-2012,BLR} indicates that superfluid
turbulence at large scales $R\gg \ell$ is similar to classical turbulence if the mechanical forcing is similar. Examples are furnished by a towed grid\,\cite{grid} forcing or by a pressure drop in a channel\,\cite{channel1,channel2}. The reason for the similarity is that the interaction of normal fluid component with the quantized-vortex tangle  leads to a mutual friction force \cite{1,2,Vinen} ``which couples together $\B u\sb n(\B r,t)$ and $\B u\sb s(\B r,t)$ so strongly that they move as one fluid"\,\cite{SR-1991}. This strong coupling effect was demonstrated analytically in \Ref{L199}  and   was later confirmed by  numerical simulations of the two-fluid model\,\cite{59,60} over a wide temperature range ($1.44<T <2.157\,$ K, corresponding to the ratio of densities $\rho\sb n/\rho\sb s$ from $0.1$ to $10$). The simulations showed strong locking of normal- and superfluid velocities at large scales, over one decade
of the inertial range. In particular, it was found that even if either the normal or the superfluid is
forced at large scale (the dominant one), both fluids get locked very efficiently. Only detailed numerical simulations (in the framework of so-called shell models of turbulence) with very large inertial interval\,\cite{PRB2015,61} showed minor decoupling of $\B u\sb s$ and $\B u\sb n$ at the viscous edge of the inertial interval in agreement with the analytical result of \Ref{L199}.

A different situation is expected for thermally driven superfluid turbulence. This type of turbulence is  generated by a heater located at the closed end of a channel which is open at the other end to a superfluid helium bath. In this case the heat flux is carried away from the heater by
the normal fluid alone with the mean velocity $\B U\sb n$, and, by conservation of mass, a superfluid current with the mean velocity $\B U\sb s$ arises in the opposite direction. This gives rise to a relative (counterflow) velocity
 \begin{equation}\label{Vns}
 \B U\sb{ns}\= \B U \sb n - \B U\sb s\,,
  \end{equation}
which is  proportional to the applied heat flux. Invariably this counterflow excites an
accompanying tangle of vortex lines.  In counterflow experiments there is no mean mass flux  and the mean velocities $\B U\sb s $ and $\B U\sb n$  of the superfluid and the  normal fluid  components are related as follows: $\rho\sb n \B U\sb n+ \rho \sb s \B U\sb s=0$.

A situation very similar to counterflow  appears in  superflows. Here superleaks (i.e. filters located at the channel end with sub-micron-sized holes permeable only to the inviscid superfluid component) allow a net flow of the superfluid
component in the channel. Contrary to counterflows, now  the normal component
remains stationary on the average: $\B U\sb n=0$.  In both counterflows and superflows the normal- and superfluid components are moving with different mean velocities and their relative velocity $U\sb {ns}\ne 0$.

Clearly, in both cases one expects properties of the normal- and superfluid velocity fluctuations different from that in the mechanically driven ``co-flow"  turbulence, in which $U\sb{n}=U\sb{s}$ and $U\sb{ns}=0$. The simple reason for that is illustrated in \Fig{F:1}, in which eddies of scales $R_1< R_2 < R_3$ are shown at three successive moments of time $t=-\tau$, $t=0$ and  $t=\tau$ for co-flow (panels  (a), (b) and (c)) and for counterflow (panels  (d), (e) and (f)).

In the co-flow the quantized-vortex tangles (shown by blue solid lines) are swept by the superfluid component with the mean velocity close to $\B U\sb s$ together with  the normal fluid eddies (shown by red dashed lines), which are swept by the normal fluid component with their mean velocity $\B U\sb n$.  Since in the co-flow $U\sb s= U\sb n$, all (normal- and superfluid eddies) are swept with the same velocity, the entire eddy configuration is moving as a whole from the left, in panel (a),  to the right  in panel (c) in the ``laboratory" reference system, shown in all panels as a black frame. During their common motion, the mutual friction effectively couples the velocities and $\B u\sb n(\B r,t)= \B u\sb s(\B r,t)$. The situation is completely different in the counter-flow, where  the mean velocities have opposite  directions and $U\sb{ns}\ne 0$. We have chosen for concreteness $U\sb n > 0 $, therefore the normal fluid (red dashed line) eddies are moving in our pictures from the left [in panel (d)] to the right [in panel (f)]. At the same time, $U\sb s < 0 $ and superfluid (blue solid line) eddies are moving in the opposite direction.

Assume that at some intermediate moment of time [chosen as $t=0$ in panel (e)] all normal- and superfluid eddies of scales $R_1$, $R_2$ and $R_3$ overlap. Choose the time-step $\tau$, such that $\tau\simeq R_2/U\sb{ns}$. The largest eddies of scale $R_3$ are almost fully overlapping during the time-step $\tau$,  while smaller eddies of scale $R_1$,  which were overlapping at $t=0$, are fully separated  at times $t\pm \tau$. Intermediate $R_2$-scale eddies are partially overlapping during the time-step $\tau\simeq \tau\sb{ol}(R_2)$. Here the ``overlapping time" of $R$-eddies  $\tau\sb{ol}(R)= R/U\sb {ns}$ is the time that is required for eddies to be swept by the counterflow velocity $U\sb{ns}$ over distance of their scale $R$.
%Therefore  the $R_2$ are partially overlapping in panels (d) and (f), while the smaller $R_1$-eddies do not overlap at the times $\pm \tau$. Similarly,  the larger $R_3$ eddies are almost overlapping for $-\tau <t< \tau$ in panels (d), (e) and (f). We see that the normal- and superfluid $R$-eddies in the counter-flow overlap in the physical space only during a limited ``overlapping" time $\tau\sb{ol}\simeq R/U\sb {ns}$.

This time may be small compared to the time $\tau\sb{cor}$ required for an effective coupling of the $\B u\sb s(\B r,t)$ and $\B u\sb n(\B r,t)$ velocities.  As we show in the last paragraph of \Sec{ss:CFE},  $\tau\sb{cor}$ is scale independent and may be estimated as $\tau\sb{cor}\sim 1/ (\kappa \C L)$, where $\C L$ is the vortex line density.  The detailed analysis shows that for most eddies in the relevant range of scales $H < R < \ell$ the time $\tau\sb{ol}\ll \tau\sb{cor}$ and therefore the velocities $\B u\sb s(\B r,t)$ and $\B u\sb n(\B r,t)$ are decoupled. This makes the energy dissipation due to mutual friction very effective and results in significant suppression of the energy spectra of the normal- and superfluid turbulent velocity spectra as compared to that in the mechanically driven turbulence, in which $U\sb {ns}=0$.

Notice that in \Ref{24} it was mentioned  that in the
counterflow, the coupling at all length scales must, to some
extent, break down, because similar eddies in the two
components are continually pulled apart, and this leads
to dissipation at all length scales.

The main goal of the present paper is to offer a relatively simple, physically transparent model of the  cross-correlation function of the normal
 and superfluid velocities, that accounts for non-zero value of the mean counterflow velocity $U\sb {ns}$.   For simplicity we consider only  the case of homogeneous and isotropic turbulence of an incompressible flow of $^4$He. In this flow the difference between the counterflow and a pure superflow turbulence disappears due to Galilean invariance. The paper is organized as follows. First we overview the two-fluid coarse-grained Hall-Vinen-Bekarevich-Khalatnikov (or HVBK)  model\,\cite{Vinen,BK61}, properly generalized for the case of counterflow turbulence, \Eqs{NSE}. Second, we suggest an approach that leads to  a crucial simplification that allows us to derive analytical equations\,\eqref{final} for the cross-correlation function of the normal- and superfluid velocity fluctuations, $\C E\sb{ns}(k,U\sb{ns})$.
Third, we analyze the equation for $\C E\sb{ns}(k,U\sb{ns})$ and show that as a rule  $\C E\sb{ns}(k,U\sb{ns})\ll  \C E\sb{ns}(k,0)$, see \Fig{F:3}.  Finally, in the concluding section, we discuss how the decoupling of velocities should affect the normal- and superfluid energy spectra.

\section{\label{s:tw-fl}Basic equations of motion for counterflow turbulence}
\subsection{Two-fluid, gradually-damped HVBK equations }
 As said above, the  large-scale motions of superfluid \He4 (with characteristic scales $R\gg \ell$) are well described by the two-fluid  model, consisting of a normal and a superfluid component with densities $\rho\sb n(T)$ and $\rho\sb s(T)$ respectively.  Neglecting both the bulk viscosity and  the thermal conductivity leads to the simplest model with two incompressible fluids, having the form of an Euler equation for  $\B u\sb s$  and a Navier-Stokes equation for   $\B u\sb n$, see, e.g.  Eqs. (2.2) and
(2.3) in  Donnely's textbook~\cite{1}. Supplemented with quantized vortices that give rise to a mutual friction force $\B F\sb{ns}$ between the superfluid and the normal components, these equations are known as  Hall-Vinen-Bekarevich-Khalatnikov (or HVBK)  model\,\cite{Vinen,BK61}:
\begin{subequations}\label{NSE-or}
\begin{eqnarray}\label{NSEa-or} %%
 \frac{\p \,\B u\sb s}{\p t}+ (\B u\sb s\* \B
\nabla)\B u\sb s   + \frac 1{\rho\sb s }\B \nabla p\sb s&=&\nu'\sb s\,  \Delta \B u\sb s -\B F \sb {ns}\,, %%
 \\ \label{NSEb-or}
  \frac{\p \,\B u\sb n}{\p t}+ (\B u\sb n\* \B
\nabla)\B u\sb n  +\frac 1{\rho\sb n }\B \nabla p\sb n&=&\nu\sb n\,  \Delta \B
u\sb n + \frac{\rho\sb s}{\rho\sb n}\B F \sb {ns}\ . ~~~~~~~
\end{eqnarray} %%
Here $p\sb n$,  $p\sb s$   are  the pressures  of the normal
and the superfluid components:
$$ p\sb n =\frac{\rho\sb n}{\rho }[p+\frac{\rho\sb s}2|\B u\sb s-\B u\sb
n|^2]\,,\
 p\sb s = \frac{\rho\sb s}{\rho }[p-\frac{\rho\sb n}2|\B u\sb s-\B u\sb
n|^2]\,,
  $$
$\rho\= \rho\sb s+\rho\sb n$  is the total density and $\nu\sb n$ is the kinematic viscosity of normal fluid. The mutual friction force is given by
$$
\B F\sb{ns}= \alpha\,  \hat {\B \omega}\times [\B \omega \times (\B u\sb{n }-\B u\sb{s})]+ \alpha' \hat {\B \omega}\times (\B u\sb{n }-\B u\sb{s})\ .
$$
In this equation $\alpha$,  $\alpha '$ are temperature dependent dimensionless mutual friction parameters and    $\B \omega$ is traditionally understood as superfluid vorticity:   $\B \omega= \B \nabla \times \B u\sb s$ and $\hat \omega \= \B \omega / |\B \omega|$.

Notice also that the original  HVBK model does not take into account the important process of vortex reconnection. In fact, vortex
  reconnections are responsible for the dissipation  of the superfluid motion due to mutual friction.

For temperatures above $1\,$K this the extra dissipation can be modeled using an effective superfluid viscosity $\nu'\sb s(T)$\,\cite{VinenNiemela}:
  \begin{equation}\label{nus}
 \nu'\sb s (T)\approx \alpha \, \kappa\ .
 \end{equation}%%
 and, following Ref.\cite{PRB2015}, we  have added  a dissipative term proportional to  $\nu'\sb s $  to the standard HVBK model.

 The effective superfluid viscosity $\nu'\sb s$  involves  a quantum-mechanical parameter $\kappa$, proportional to the Plank's constant $h$. This underlies the fact that the corresponding term in \Eqs{NSE-or} originates from the motions of quantized vortex lines at quantum scales $\sim \ell$. This is not captured by the coarse-grained, classical HVBK equations.

 Bearing in mind that experimentally the counterflow cannot be realized for $T<1\,$K (due to practically zero normal fluid density) we cannot discuss here the delicate issue how to account for the superfluid dissipation in \Eqs{NSE-or} for such low temperatures.
\end{subequations}

\subsection{\label{ss:CFE}Counterflow HVBK equations }

\begin{figure}
  \ \includegraphics[width=9cm, keepaspectratio]{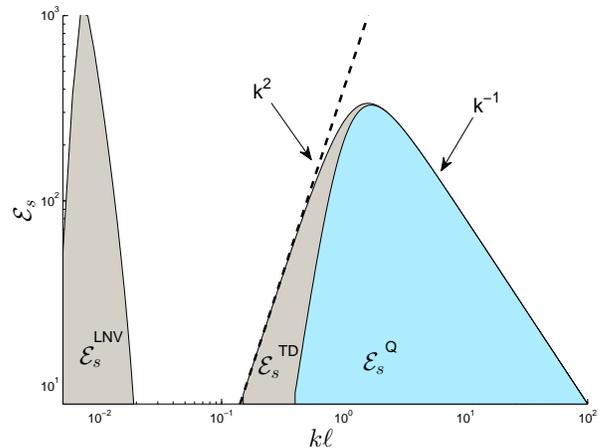}

   \caption{\label{F:2} Color online. The sketch of the stationary superfluid turbulent energy spectrum in the counterflow [log-log coordinates,  $ \log \C E\sb s(k)$ vs. $\log (k\ell)$]. The spectrum  $  \C E\sb s(k)$ consists of a classical   $ \C E\sp {cl}\sb s(k)$ and a quantum $ \C E\sp {qn}\sb s(k)$ parts, colored in gray  and light blue, respectively.  For concreteness, as a large-scale classical peak we used here Lvov-Nazarenko-Volovik spectrum\,\eqref{LNV}, found for $^3$He with resting normal fluid component, but presumably valid for counterflowing $^4$He in the $k$-range with fully decoupled the normal- and superfluid velocities. The quantum (light blue) contribution $  \C E\Sp Q\sb s(k)$ has  $1/k$ asymptotics at large k, originated from superfluid motions near the vortex cores. It is adjacent to the classical thermal bath part
 $ \C E\Sp {TD}\sb s(k)\propto k^2$ with equipartition of energy between degrees of freedoms.}
\end{figure}

 To proceed we separate the mean velocities $\B U\sb n$ and $\B U\sb s$ from  the turbulent velocity fluctuations, $\B u'\sb n(\B r,t)$ and $\B u'\sb s(\B r,t)$ with zero mean. Equations\,\eq{NSE-or} for $\B u'\sb n(\B r,t)$ and $\B u'\sb s(\B r,t)$ may be written,  as follows:
\begin{subequations}\label{NSE}
\begin{eqnarray}\label{NSEa} %%
 \Big (\frac{\p \,}{\p t}+ \B U\sb s \cdot \B \nabla  -\nu'\sb s\,  \Delta \Big )\B u'\sb s +\mbox{NL}\{\B u'\sb s,\B u'\sb s\}     &=&   -\B f' \sb {ns}\,,~~~ %%
 \\
  \Big (\frac{\p \,}{\p t}+ \B U\sb n \cdot \B \nabla  -\nu\sb n\,  \Delta \Big )\B u'\sb n +\mbox{NL}\{\B u'\sb n,\B u'\sb n\}     &=&   \frac{\rho\sb s}{\rho\sb n}\B f' \sb {ns} . ~~~~~\label{NSEb}
   \end{eqnarray}
   Here the nonlinear terms NL$\{\B u'\sb s,\B u'\sb s\} $ and NL$\{\B u'\sb n,\B u'\sb n\} $ are quadratic in the corresponding velocities functionals.  These terms originate from the terms $\B  u'\cdot \B \nabla \B  u'$   and from the  $\B \nabla  p'$ terms, where the pressure  fluctuations $p'(\B r,t)$ were expressed via a quadratic velocity fluctuations functional, using the incompressibility condition. For our purpose we will not need to specify the nonlinear terms  NL$\{\B u'\sb s,\B u'\sb s\} $ and NL$\{\B u'\sb n,\B u'\sb n\} $.

 Next we approximate the mutual friction fluctuation term $\B f'\sb{ns}$. In the spirit of  Ref.~\cite{LNV}, we write as follows: %%
\begin{equation}\label{NSEf-a}
  \B f'\sb {ns}\simeq -\a(T)\,(\B  u'\sb n-\B  u'\sb s ) \,\Omega \ . \end{equation}

 In \Ref{LNV} the characteristic superfluid vorticity $\Omega  $ in \Eq{NSEf-a} was understood as the root-mean-square (rms) vorticity: $\Omega\simeq \sqrt {\< |\B \omega|^2\> }$. However in counterflow turbulence there is an additional quantum mechanism of creating vortex lines, elucidated
in pioneering works by Schwarz
\cite{19}:  the force of mutual friction can lead to the stretching
of the vortex lines, and this in turn can lead to a
self-sustaining turbulence in the superfluid component
provided that vortex lines are allowed to reconnect. This mechanism is leading to the creation of an additional peak in the superfluid energy spectrum near the intervortex scale $\ell$,  sketched in \Fig{F:2}.  In  the counterflowing superfluid turbulence this peak provides the main contribution to the rms vorticity, which cannot be described in the framework of the coarse-grained HVBK \Eqs{NSEa} and \eqref{NSEb}, which is valid only for scales $R\gg \ell$. Therefore  $\Omega$ in \Eq{NSEf-a} should be understood as an external parameter in the HVBK equations for the counterflow, simply estimated via the vortex line density $\C L$, which in its turn is proportional to the square of the counterflow velocity:
\begin{equation}\label{Omega}
\Omega \simeq \kappa \C L\,, \quad \C L\approx (\gamma_{_{\C L}} U\sb {ns})^2\ .
\end{equation}
\end{subequations}
 Here $\gamma_{_{\C L}}$ is a temperature dependent  phenomenological parameter that  varies from about  70 s/cm$^2$ to about  150 s/cm$^2$ when $T$ grows from 1.3\,K to 1.9\,K (see e.g.   Fig 9 in \Ref{BLPP-2013}) We have added here a subscript ``$_{_{\C L}}$" to distinguish the traditional notation $\gamma$ in \Eq{Omega} from the characteristic frequencies $\gamma\sb s$ and $\gamma \sb n$ that are used below.
%--------------------------------------------
%--------------------------------------------

  The resulting gradually damped HVBK model for turbulent counterflow in $^4$He,  \Eqs{NSE},  serves as a basis for our study of the correlations between normal- and superfluid velocity correlations. We will refer to these equations as the ``counterflow HVBK equations".

  Equations\,\eqref{NSE} allow to estimate  the  time $\tau\sb{cor}$ required for the coupling of the  normal and superfluid turbulent velocities by mutual friction. To this end we consider an equation for their difference, $\B u'\sb{ns}\= \B u'\sb{n}-\B u'\sb{s}$, subtracting \Eq{NSEa} from \Eq{NSEb}:
  $$
 \frac{\p \,\B u'\sb{ns}}{\p t}+  \dots    = - \, (\kappa \C L)  \, \alpha \sb{ns} \B u'\sb{ns}\,, \quad \alpha \sb{ns}\=  \frac {\alpha \rho }{\rho\sb n}\ .~~~ %%
$$
  Here we dempted by $\dots$ the sweeping, viscous and nonlinear terms that are irrelevant for the current discussion. Evidently,  $\tau\sb{cor}$   should be estimated as $1/ ( \alpha\sb {ns}\kappa \C L)$. The  temperature dependence of $\alpha\sb{ns}$,  shown in \Fig{F:4} by a red line with squares,  indicates that $\alpha\sb {ns}\sim 1$.  Therefore we can conclude that $\tau\sb{cor}\sim 1 / ( \kappa \C L)$, as mentioned in Sect. \ref{s:intro}.

 %==================================================
 \section{\label{s:corr}Normal - superfluid velocity correlations in    $^4$He}
 The main result of this Section is \Eq{final} for the cross-correlation function  of the normal- and superfluid velocity turbulent fluctuations in a stationary, space homogenous counterflow $^4$He-turbulence. This equation   describes how the cross-correlations depends on the counterflow velocity, the scale (wave-number) and the temperature. Its derivation requires some definitions and relationships that are common in statistical physics. We recall them in Appendix A.

\subsection{\label{ss:derivation} Derivation of the cross-correlation ${\C E\sb{ns}( k)}$}

The first step in the derivation of the cross correlation is rewriting the
counterflow HVBK  \Eqs{NSE}  in $(\B k,t)$-representation, defined by \Eq{FTa}:
\begin{subequations}\label{NSEkt}\begin{eqnarray}\nn  %%
&&  \Big (\frac{\p \,}{\p t}+ i \B U\sb s \cdot \B k  +\nu'\sb s k^2 + \Omega\sb s \Big ){\B v }\sb s \\ \label{NSEktA}
&+&\mbox{NL}_{\B k}\{{\B v }\sb s,{\B v }\sb s\}=   \Omega \sb s {\B v }\sb n\, ,~~~ \\ \nn %%
 &&  \Big (\frac{\p \,}{\p t}+ i \B U\sb n \cdot \B k  +\nu \sb n k^2 + \Omega\sb n \Big ){\B v }\sb n \\  \label{NSEktB}
&+&\mbox{NL}_{\B k}\{{\B v }\sb n,{\B v }\sb n\}=   \Omega \sb n {\B v }\sb s\, ,~~~ \end{eqnarray}
where the mutual friction frequencies are given by
\begin{equation}\label{NSEktC}
 \Omega\sb s\= \alpha\, \Omega\,, \quad \Omega\sb n\=  \alpha\sb n \, \Omega\,,  \alpha\sb n  \= \alpha \, \rho\sb s/\rho \sb n \ .
\end{equation}
     \end{subequations}
The nonlinear terms $ \mbox{NL}_{\B k}\{{\B v }\sb s,{\B v }\sb s\}$ and $\mbox{NL}_{\B k}\{{\B v }\sb n,{\B v }\sb n\}$ in \Eqs{NSEktA} and \eqref{NSEktB} couple all $\B k$-Fourier harmonics making their analytic solution intractable. To proceed we therefore simplify the equations in the spirit of the Direct Interaction Approximation (DIA) that was developed by Kraichnan for classical turbulence\,\cite{Kra}. This approximation is equivalent to a 1-loop truncation of the  Wyld  diagrammatic expansion\,\cite{Wyld} of the nonlinear equations with a 1-pole approximation \cite{acoustic} for the Green's function. While uncontrolled, this approximation served usefully in the study of classical turbulence, and we propose that it is also useful in the present context.  The upshot of the DIA approximation is a rewriting of the nonlinear terms in \Eqs{NSEktA} and \eqref{NSEktB} as a sum of two contributions\,\cite{LP}:
\begin{subequations}\label{approx}
\begin{eqnarray}\label{approxA}
\mbox{NL}_{\B k}\{{\B v }\sb s,{\B v }\sb s\}&=& \gamma\sb s (k)\B v \sb s(\B k,t)- \B \varphi\sb s(\B k,t)\,,\\ \label{approxB}
\mbox{NL}_{\B k}\{{\B v }\sb n,{\B v }\sb n\}&=& \gamma\sb n (k)\B v \sb n(\B k,t)- \B \varphi\sb n(\B k,t)\ .
\end{eqnarray}
The  $\gamma\sb s (k)$ and $\gamma\sb n (k)$ are the charateristic frequencies and $\B \varphi\sb s(\B k,t)$ and $\B \varphi\sb n(\B k,t)$ are the force terms.
The  terms proportional to $\gamma\sb s (k)$ and $\gamma\sb n (k)$ describe the energy flux from fluctuations with given $\B k$ to all others degrees of freedom.  In classical turbulence theory these characteristic frequencies are referred to as ``turbulent viscosity" and estimated as follows:
\begin{equation}\label{NSE2e}
   \gamma\sb n (k)\simeq \sqrt {k^3 \C E\sb n (k)}\,, \quad  \gamma\sb s (k)\simeq \sqrt {k^3 \C E\sb s  (k)}\ .
\end{equation}
 In turbulent systems with strong interactions these frequencies  are the inverse turnover times of eddies of scale $1/k$.

The force terms in the approximation\,\eqref{approxA} and \eqref{approxB} mimic the energy influx  to  fluctuations with given $\B k$ from all others degrees of freedom. In the simplest Langevin approach these forces are random Gaussian processes  with zero mean and $\delta$-correlated in time:
\begin{eqnarray} \nn
\<\B \varphi\sb s (\B k, t)\cdot \B \varphi\sb s^*  (\B k', t') \>&=& (2\pi)^3 \d (\B k -\B k') \delta(t-t') \varphi^2\sb{ss}(\B k) \,, \br
\<\B \varphi\sb n (\B k, t)\cdot \B \varphi\sb n^*  (\B k', t') \>&=& (2\pi)^3 \d (\B k -\B k') \delta(t-t')\varphi^2\sb{nn}(\B k) \,, \\
\label{FF}
\<\B \varphi\sb s (\B k, t)\cdot \B \varphi\sb n^*  (\B k', t') \> &=& 0\ .
\end{eqnarray}
\end{subequations}

 Here the Delta functions $\delta(\B k-\B k')$ originate from the space homogeneity. An important difference from the  traditional Langevin approach is that our turbulent system is not in the thermodynamic equilibrium and therefore  the correlation amplitudes $\varphi^2\sb{nn}$ and $\varphi^2\sb{ss}$ are not determined by fluctuation-dissipation theorems.  We will show below that these amplitudes may be expressed via the energy spectra $\C E\sb s(k)$ and  $\C E\sb n(k)$.

With these approximations the counterflow HVBK \Eqs{NSEkt} become linear in $\B v\sb s$ and $\B v\sb n$:
\begin{subequations}\label{NSE2}\begin{eqnarray}\label{NSE2a} %%
&& \hskip -.5cm  \Big [ \frac{\p   }{\p t}+ i \B k \cdot \B U\sb s + \G \sb s \Big ]   \B v\sb s(\B k, t) =\Omega  \sb s  \B v\sb n(\B k, t)  + \B \varphi\sb s(\B
  k, t),  ~~~~~~~ %%
 \\ \label{NSE2b}
&& \hskip -.5cm  \Big [ \frac{\p   }{\p t}+ i \B k \cdot \B U\sb n + \G \sb n \Big ]   \B v\sb n(\B k, t)  = \Omega  \sb n  \B v\sb s
 (\B k, t) + \B \varphi\sb n(\B k, t), \\   \label{NSE2c}
  &&  \G \sb n=\g\sb n +\Omega \sb n    + \nu\sb n k^2\,, \quad  ~\G \sb s=\g \sb s + \Omega  \sb s+  \nu'\sb s k^2\ . ~~~~ \label{NSE2d}
   \ .
\end{eqnarray}\end{subequations}
Clearly, counterflow turbulence in a channel is anisotropic due to the existence of two preferred directions: the stream-wise direction $\B x$ and the wall-normal direction $\B y$. Even far away from the wall, in the channel core, where classical hydrodynamic turbulence can be treated as isotropic, in quantum turbulence there remains one preferred direction $\B x$ of the counterflow velocity $\B U\sb{ns}$. Schwarz\,\cite{19} introduced an anisotropy index $I_\parallel $, equal to $2/3$ in the case of isotropy. Numerical simulations (see, e.g. \Ref{Kond})  shows that $I_\parallel$ varies between 0.74 and 0.82, depending on the temperature and the counterflow velocity. Therefore the dimensionless measure of anisotropy $3 I_\parallel/2 -1$ is below 20\% in any case. According to our understanding, this level of anisotropy cannot affect significantly the results presented below. Aiming at simplicity and transparency of the derivation we assume isotropy from the very beginning, leaving a more general derivation (in the framework of the same formal scheme) for the future. For weak anisotropy all our results should be understood as angular averages.

Multiplying \Eqs{NSE2a} and \eq{NSE2b} by $\B v\sb s^* $, and  $\B v\sb n^* $,  respectively  and averaging, we get equations for the velocity  correlations $E\sb{nn}$, $E\sb{ss}$ and the cross-correlation $E\sb{ns}$, defined by \Eqs{def6}:
\begin{subequations}\label{CF}\begin{eqnarray}\label{CFa} %%
&&  \Big[\frac{\p  }{2\,\p t}+   \G \sb s \Big ] E\sb {ss}   =   \Omega \sb s  \mbox{Re}[E\sb {ns}]  +   \mbox{Re} \big[ \Phi\sb
  {ss}\big ]\,,  ~~~~~~~~%%
 \\ \label{CFb}
 && \Big[\frac{\p  }{2\, \p t}+   \G \sb n \Big ] E\sb {nn}   =   \Omega \sb n \mbox{Re}[E\sb {ns}] +   \mbox{Re}
 \big[ \Phi\sb {nn}\big ]\,,  ~~~~~~~~%%
 \\ \nn
&&  \Big[\frac{\p  }{\p t}+ i \B k \cdot \B U\sb {ns}+ \G \sb s + \G \sb n \Big ] E\sb {ns}   = \Big[\Omega \sb s   E\sb {nn} +
 \Omega \sb n  E\sb {ss}\Big] \\  \label{CFc}
  && \hskip 4.9cm   + \, \Phi^* \sb {sn}+ \Phi\sb {ns} \ .  ~~~~~~~~%%
 \end{eqnarray}\end{subequations}
These equations involve the presently unknown simultaneous cross-correlations   of the velocities and the forces, $\Phi_{\dots}$,  defined similarly to
\Eqs{corr}: %%%
\begin{subequations}\label{corr1}%%
  \begin{eqnarray}\label{corr-nn}%%
    \< \B \varphi\sb n(\B k,t)\*\B v^* \sb n(\B k',t)  \>&=& (2\pi)^3 \Phi\sb{nn}(\B k)\, \d(\B k-\B k')\,, \\ %%
    \label{corr-ss}%%
    \< \B \varphi\sb s(\B k,t)\*\B v^* \sb s(\B k',t)  \>&=& (2\pi)^3 \Phi\sb{ss}(\B k)\, \d(\B k-\B k')\,, \\ %%
    \label{corr-ns}%%
    \< \B \varphi\sb n(\B k,t)\*\B v^* \sb s(\B k',t)  \>&=& (2\pi)^3 \Phi\sb{ns}(\B k)\, \d(\B k-\B k')\,, \\ %%
    \< \B \varphi\sb s(\B k,t)\*\B v^* \sb n(\B k',t)  \>&=& (2\pi)^3 \Phi\sb{sn}(\B k)\, \d(\B k-\B k')\ .~~~~~
\end{eqnarray}\end{subequations}
To find these correlations, we rewrite  \Eqs{NSE2} in Fourier ($\B k, \o$)-representation:
  \begin{subequations}\label{NSE3}\begin{eqnarray}\label{NSE3a} %%
 \big [i(\B k \cdot \B U\sb s-\o) & +& \G \sb s  \big ]  \~{\B v} \sb s(\B k, \o) \\ \nn &=&\Omega \sb s \~{\B v}\sb n(\B k, \o)  + \~{\B \varphi}\sb s(\B k,
 \o)\,,  ~~~~~~~~%%
 \\ \label{NSE3b}
 \big [i( \B k \cdot \B U\sb n-\o) &+ &\G \sb n  \big ]   \~{\B v}\sb n(\B k, \o) \\ \nn & =& \Omega \sb n  \~{\B v}\sb s (\B k, \o) +
  \~{\B \varphi} \sb n(\B k, \o)\ ,
 \end{eqnarray}\end{subequations}
 where $\~{\B \varphi}\sb s(\B k,\o)$ and $\~{\B \varphi}\sb n(\B k,\o)$ are the ($\B k,\o$)-representation of the force terms $\B \varphi\sb s (\B k, t)$ or $\B \varphi\sb
n (\B k, t)$.
 The solution of the linear \Eqs{NSE3} reads:
\begin{subequations}\label{sol3}\begin{eqnarray}\label{sol3a}
  \~{\B v} \sb s  &=& -  \big[ (i(  \B k \cdot \B U\sb n -\o) + \G \sb n) \~ {\B \varphi}\sb s    + \Omega \sb s  \~ {\B \varphi}\sb n \big ]/ \D  \,, \\
 \~{\B v} \sb n  &=& -  \big[ (i(  \B k \cdot \B U\sb s -\o) + \G \sb s) \~ {\B \varphi}\sb n    + \Omega \sb n  \~ {\B \varphi}\sb s \big ]/ \D  \,,\\ \label{sol3c}
 \D &\=& (\omega -\B k\cdot \B U\sb n+i \Gamma\sb n)(\omega -\B k\cdot \B U\sb s+i \Gamma\sb s)\\ \nn &&+\O \sb n \O\sb s \,,
 \end{eqnarray}\end{subequations}
where for brevity we suppressed the arguments $(\B k, \o)$ in all functions.

Multiplying the two \Eqs{sol3} by $\~{\B \varphi} \sb n$ and $\~{\B \varphi}
\sb s$,  respectively  and averaging, we get equations for the (cross)-correlations   $\~ \Phi\sb{ns}(\B k,\o)$ and $\~ \Phi\sb{sn}(\B k,\o)$  which give after  integration  over $\o$   the simultaneous cross-correlation functions:
\begin{subequations}\label{Fns}
\begin{eqnarray}
 \Phi\sb{sn}(\B k) &=&   -\frac{\Omega \sb s f^2 \sb n}{2\pi}\int  \frac {d\omega} {\Delta^*(\B k,\o)}=0\,, \\
  \Phi\sb{ns}(\B k)&=&  -\frac{\Omega \sb n f^2 \sb s}{2\pi}\int  \frac {d\omega} {\Delta^*(\B k,\o)}=0 \ .
\end{eqnarray}\end{subequations}
To compute the above integrals we found the solutions of the equations $\Delta (\B k,\o)=0$ with respect to $\omega$:
\begin{eqnarray}\label{om}
\omega&=&\omega_\pm= \frac i 2 \Big [ -i \B k\cdot (\B U \sb n+ \B U \sb s) + \Gamma\sb s+ \Gamma\sb n \Big]\\ \nn
&& \pm \sqrt{\big(\Gamma\sb s- \Gamma\sb n+ i \B k \cdot \B U\sb {ns}\big )^2+ 4 \O\sb s \O\sb n} \ .
\end{eqnarray}
Using these solutions, after relatively simple analysis, we find that both roots have positive imaginary parts: Im$[\o_+]>0$ and Im$[\o_-]>0$. Therefore, the integral in \Eqs{Fns} vanishes. Now \Eq{CFc} in the stationary case gives:
\begin{subequations}\label{res4} \begin{eqnarray}\label{res4a}
E\sb{ns}(\B k)&=& \frac{A}{B+i \B k \cdot \B U \sb{ns} }\,,\\
 A&\=&\Omega\sb s E\sb {nn}(k)+\Omega\sb n E\sb {ss}(k)\,, \\
  B&\=&\Gamma\sb n + \Gamma \sb s \ .
\end{eqnarray}
Averaging \Eq{res4a} with respect to all orientations of $\B k$ we get:

\begin{equation}\label{res4d}
 \< E\sb{ns}(\B k)\>\sb{angle}= \frac A {k U\sb {ns}} \arctan  \Big (\frac   {k U \sb {ns}} B \Big )\ .
\end{equation}
\end{subequations}
Using \Eqs{def1a} this can be finally rewritten as follows:
\begin{subequations}\label{final}
\begin{eqnarray}\label{finalA}
\C E \sb {ns}(k) &=& \C E ^{(0)}\sb {ns}(k)D(\zeta)\,,
 \\ \label{finalB}
\C E ^{(0)}\sb {ns}(k) &=& \frac {\Omega \sb s \C E \sb{n}(k)+ \Omega \sb n \C E \sb{s}(k)} {\Gamma\sb {s}(k) +\Gamma \sb n(k)} \,,\\ \label{finalC}
D(\zeta)&=&\frac 1{\zeta(k)}\arctan [\zeta(k)]\,, \\ \label{finalD}
\zeta(k)&\=& \frac{k U\sb {ns}}{  \Gamma\sb s (k)+ \Gamma \sb n (k)}\ .
\end{eqnarray}
Here $\C E^{(0)} \sb {ns}(k)$ is the cross-correlation function for zero counterflow velocity which was previously found in \Ref{L199}. The dimensionless ``decoupling function" $D(\zeta)$ of the dimensionless ``decoupling parameter" $\zeta(k)$, describes  the decoupling of the normal- and superfluid velocity fluctuations, caused by the counterflow velocity.

Notice that in future comparisons of the experimental or numerical data with \Eqs{final} one needs to bear in mind that the counterflow velocity affects not only the decoupling function $D(\xi)$, but also the energy spectra $\C E\sb n(k)$ and $\C E\sb s(k)$ in \Eq{finalB} for $\C E\sb{ns}^{(0)}(k)$.

%\green{A preliminary discussion of this effect can be found in Subsec. III C 1 of \Ref{Prague}}
.
\end{subequations}
\begin{subequations}\label{limits}

Considering the limits of small and large values of the decoupling parameter $\zeta$ we get from \Eq{finalA}:
\begin{eqnarray}\label{limitsA}
\C E \sb {ns}(k) &=& \Big [ 1- \frac {\zeta (k)^2}3\, \Big ]\C E^{(0)} \sb {ns}(k)\,, \ \mbox{for}\  \zeta(k)\ll 1\,, ~~~~~~\\ \label{limitsB}
\C E \sb {ns}(k)&=& \frac{\pi}{2\, \zeta(k)}\, \C E^{(0)} \sb {ns}(k)\,, \quad ~~~~~\,\mbox{for}\  \zeta(k)\gg 1\ .
\end{eqnarray}
\end{subequations}

\begin{figure}%[h!]

  \includegraphics[width=\columnwidth, keepaspectratio]{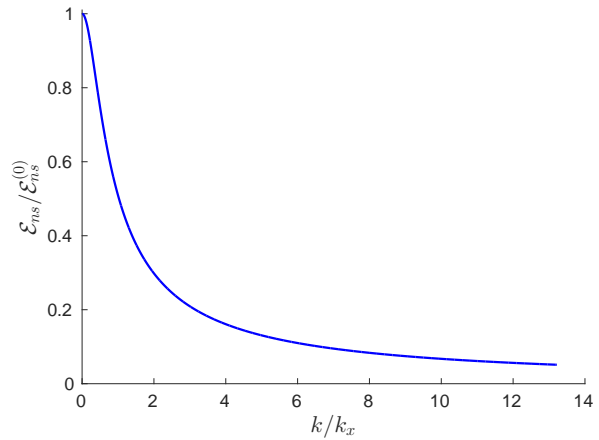}
  \caption{\label{F:3} The decoupling function $\C E\sb {ns}/ \C E\sb {ns}^{(0)}=D(k/k_\times) $  vs the dimensionless wave number $k/k_\times$.}
\end{figure}

We choose the crossover value  $\zeta_\times\approx 2$  such that $D(\zeta_\times)=1/2$. Below we show that with  good accuracy $\zeta(k)\propto k$.  Therefore, we can consider $D[\zeta(k)]$ as a function of $k$ and present in Fig. \ref{F:3} the decoupling  ratio due to counterflow velocity $\C E\sb {ns}(k)/ \C E^{(0)}\sb {ns}(k)=D[\zeta(k)]$ , as a function of $k/k_\times$. Our estimate below shows that the crossover wave number $k_\times$ (for which $\C E\sb{ns}(k) = \C E\sb{ns}^{(0)}(k)/2$) is independent  of the counterflow velocity and typically  is in the relevant interval of scales, between  $\pi/H$ and $\pi/\ell$.

 \begin{figure}%[h!]

  \includegraphics[width=\columnwidth, keepaspectratio]{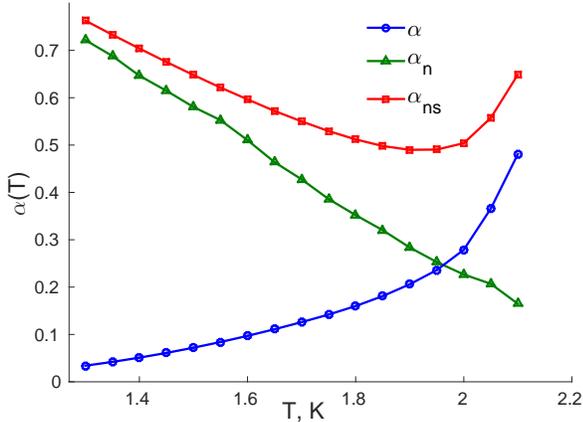}
  \caption{\label{F:4} Color online. Temperature dependence of the mutual friction parameters  for $^4$He, \Ref{DonnelyBarenghi98}: $\alpha$ for the superfluid \Eq{NSEa}, blue   line with   circles;  $ \alpha\sb n (T)=\alpha \rho\sb s /\rho\sb n$  in the normal fluid  \Eq{NSEb}, green   line with   triangles; and $ \alpha\sb {ns}=\alpha+ \alpha\sb n= \alpha \rho  /\rho\sb n$ in \Eqs{Gammas}, red line with squares.}
\end{figure}
%------------------------------------------------------------

%===========================================
\subsection{Typical value of  the decoupling parameter $\zeta(k)$}

To clarify what are the typical values of $\zeta(k)$ in realistic conditions and how $\zeta(k)$ depends on the temperature and the counterflow velocity we note\,\cite{BLPP-2013} that the main contributions to $\Gamma\sb n$ and $\Gamma\sb s$, \Eq{NSE2c}, come from $\Omega\sb n$ and $\Omega\sb s$, given by \Eq{NSE2d}:
\begin{equation}\label{Gammas}
 \Gamma\sb n +\Gamma\sb s\approx  \Omega\sb n+\Omega\sb s = \alpha \sb {ns}\, \Omega\,,  \  \alpha \sb {ns}=\alpha+\alpha\sb n= \frac{\alpha\,\rho}{\rho\sb n} \ .
\end{equation}
Indeed, for scales $k\ell \ll 1$ the viscous terms $\nu\sb{s,n}k^2\gg \gamma\sb{s,n}(k)$ and may be safely neglected, while for scales near the intervortex distance they are of the same order of magnitude. Moreover, for  $k\ell\sim  1$, $\nu\sb{s,n}k^2\sim \gamma\sb{s,n}(k)\sim \Omega\sb{s,n}$, if one estimates $ \Omega\sb{s,n}\sim \Omega\sb{cl}$ in a classical manner via the root-mean square of the  vorticity, see e.g. \Refs{LNV,245}: $ \Omega\sb{cl}\sim \sqrt {\< \omega^2\>}$. However, as we explained above, in the counterflow there is an additional quantum mechanism of the random vortex tangle excitation with scales if the order of $\ell$. This mechanism provides the leading contribution to  $ \Omega\sb{s,n}$ and, consequently, the leading contribution to  $ \Gamma\sb{s,n}$, as written in \Eq{Gammas}.

The temperature dependence of $\alpha\sb {ns}(T)  =B(T)/2$, where $B(T)$ is the coefficient in the Vinen equation, tabulated in \Ref{DonnelyBarenghi98}, is shown in \Fig{F:4} together with $\alpha(T)$ and $\alpha\sb n(T)$.
 The  opposite temperature dependence of $\alpha (T)$ and $\alpha\sb n(T)$ results in a weak temperature dependence of the parameter $\alpha\sb{ns}(T)$ in \Eq{Gammas}; it  varies  between 0.7 and 0.5 in the relevant for counterflow experiments temperature range $1.4\div 1.9\,$K  .

Now   \Eqs{Omega} and  \eq{Gammas} together with \Eq{finalD} give:
\begin{subequations}\label{zetas}
\begin{equation}\label{zetasA}
\zeta(k)\simeq \frac{k}{\alpha\sb{ns} \,\kappa \, \gamma_{_{\C L}}^2 \, U\sb {ns}}\ .
\end{equation}
Clearly,
$\zeta(k)\propto k$ and it reaches its maximal value $\zeta\sb{max}$ at the highest $k$ value which is
 permissible in our approach, i.. $k\simeq k\sb{max}\simeq \pi /\ell$; this is  at the edge of the applicability. With $\ell\simeq 1/\sqrt {\C L}\simeq 1/ (\gamma_{_{\C L}} U\sb {ns})$ this gives a  simple estimate of $\zeta\sb{max}$, independent of $U\sb {ns}$:
 \begin{equation}\label{zetasB}
 \zeta\sb{max}\simeq \frac \pi {\alpha\sb{ns}\, \kappa \, \gamma_{_{\C L}}}\sim 50\,,  \quad \mbox{for}\  T\approx 1.4\,\mbox{K}\ .
\end{equation}
 Here for the numerical estimate we used  $\alpha\sb {ns} \simeq 0.6$,  $\gamma_{_{\C L}} \simeq 100 $ s/cm$^2$ and $\kappa \approx 10^{-3}$cm$^2$/s.   An important conclusion is that for large $k$ the normal- and superfluid velocities are practically fully decoupled: for $k\sim k\sb{max}$ $\zeta(k)\sim 50$ and
the ratio $\C E\sb{ns}/\C E\sb{ns}^{(0)}$ is about 0.03 according to \Eq{limitsB}.

An even more important conclusion is that according to \Eq{zetasB} the range of wave numbers $k\sb {max}>k> k_\times$, where  $\C E\sb{ns}/\C E\sb{ns}^{(0)}<1/2$, extends over more than one decade:
\begin{equation}\label{zetasC}
\frac{\zeta\sb {max}}{\zeta_\times} \simeq\frac{k\sb {max}}{k_\times} \simeq\frac \pi {2  \alpha\sb{ns}\, \kappa \, \gamma_{_{\C L}}}\sim 25\,, \quad \mbox{for}\  T\approx 1.4\,\mbox{K}\ .
\end{equation}

Equation\,\eqref{zetasA} allows us to estimate also the minimal value
$\zeta\sb{min}$,  which is attained at $k\sb{min}\simeq   \pi / H$:
 \begin{eqnarray}
\zeta\sb{min}&\simeq& \frac{ \pi}{H  \alpha\sb{ns} \kappa \gamma_{_{\C L}}^2 U \sb {ns}} \sim  0.5\,,
\\ \nn
&&  \mbox{for} \ T\approx 1.4\,\mbox{K}, \   U\sb{ns}=1\frac{\mbox {cm}}{\mbox{s}}\,, \ H = 1~ \mbox{cm}\ .
\end{eqnarray}
This means that the value $k_\times$, for which $\zeta(k_\times)=2$, is few times larger than $k\sb{min}\simeq \pi/H$. Therefore
 for large scales (between $H$ and $R_\times \simeq \pi /k_\times$) we expect significant coupling  of the normal- and
superfluid velocities: for $\zeta=0.5$ \Eq{limitsB} give  $\C E\sb{ns}/\C E\sb{ns}^{(0)}\simeq 0.9$. The value of $\zeta\sb{min}$ is inversely proportional to $U\sb {ns}$ and for $U\sb{ns}> 1 \,$cm/s become
  even smaller than 0.5. Accordingly, for  $U\sb{ns}> 1 \,$cm/s the interval between $H$ and $R_\times$ become larger and the coupling between the normal and superfluid velocities at the largest scale $H$ is even stronger: the ratio  $\C E\sb{ns}/\C E\sb{ns}^{(0)}> 0.9$.\\
\end{subequations}
%========================================================
\section{Summary and discussion}   We demonstrated that the cross-correlation function between normal- and superfluid  velocity fluctuations $\C E\sb{ns}(k)$ in a turbulent counterflow  of $^4$He is  strongly affected by the relative  velocity $U\sb{ns}$. As described by \Eqs{final} and illustrated in \Fig{F:3}, this effect is governed by a dimensionless decoupling parameter $\zeta(k)\propto k / U\sb{ns}$, given by \Eq{zetasA}. This parameter increases with $k$ and when $k \simeq k\sb{max}\simeq \pi/\ell$ it reaches its maximum $\zeta\sb{max}\gg 1$, as estimated in \Eq{zetasB}. Accordingly,  the normal- and superfluid velocity fluctuations of small scales (i.e. for large wave numbers) are almost fully decoupled: the correlation $\C E\sb {ns}$ is much smaller than its value $\C E\sb {ns}^{(0)}\approx 1$ for $U\sb{ns}=0$. On the contrary, at large scales the energy containing fluctuations of  $R\sim H$ are almost fully coupled: $\C E\sb {ns}^{(0)}-\C E\sb {ns}\ll \C E\sb {ns}^{(0)}$. The  crossover scale $R_\times$, for which $\C E\sb {ns}=\frac 12\, \C E\sb {ns}^{(0)}$ is a few times smaller  than $H$. Therefore the large scale fluctuations, for $H \gtrsim R \gtrsim R_\times$  may be qualitatively considered as coupled:  $\C E\sb {ns} \geqslant \frac 12\C E\sb {ns}^{(0)}$. On the other hand, in the large interval of small scales, for $ R_\times \gtrsim R\gtrsim \ell$ the normal- and superfluid velocities   may  be considered as effectively decoupled:  $\C E\sb {ns}  \leqslant\frac 12\C E\sb {ns}^{(0)}$.

The coupling or decoupling of normal- and superfluid velocities crucially affects the energy dissipation due to the mutual friction. Correspondingly it also affects  the energy spectra $\C E\sb s(k,t)$ and $\C E\sb n(k,t)$. To see this let us consider the evolution equations for these objects, which may be obtained multiplying \Eq{NSEa} and \eqref{NSEb} in $(\B k, t)$-representation by $\B v\sb s(\B k,t)$    and $\B v\sb n(\B k,t)$ respectively, and averaging with respect to the turbulent statistics and directions of $\B k$ :
\begin{subequations}\label{NSE1}
\begin{eqnarray}\label{NSE1s}
 \left[\frac{\partial }{2 \,\partial t}+k^2 \nu' \sb s \right]\C E\sb s(k,t)&\!+&\! \C {N\! L}\sb s   \\ \nn
  &=& \Omega\sb s \big [ \C E\sb{ns}(k,t)- \C E \sb s (k,t) \big ]\,,~~~~~
\\ \label{NSE1n}
  \left[\frac{\partial }{2 \,\partial t}+k^2 \nu\sb n \right]\C E\sb n(k,t) &\!+&\! \C {N\! L}\sb n  \\ \nn
  &=& \Omega\sb n \big [ \C E\sb{ns}(k,t)- \C E \sb n (k,t) \big ]\ .~
\end{eqnarray}\end{subequations}
Here $\C {N\! L} \sb{s,n}$ are nonlinear terms. For   $k\gg   k_\times$, due to the decoupling $\C E\sb {ns}(k)\ll \C E\sb {s}(k)$. Therefore it may be neglected  on the RHS of \Eq{NSE1s}, which becomes $-\Omega\sb s \C E\sb s$. This is similar to the equation for $\C E\sb s$ for superfluid turbulence in $^3$He, where mutual friction drastically suppresses the energy spectrum $\C E\sb s(k)$\,\cite{L199,LNV,Sabra1}; instead of the classical Kolmogorov spectrum $\C E(k)\propto k^{-5/3}$ one finds the spectrum discussed by Lvov, Nazarenko and Volovik\,\cite{LNV}:
  \begin{equation}\label{LNV}
  \C E\sb s(k)\propto \frac{1}{k^{5/3}} \Big [\frac 1 {k^{2/3}}- \frac 1 {k^{2/3}_*} \Big ]^2\,,
   \end{equation}
   that terminates at some critical value $k_*$.
  This means that, provided that there exists a full decoupling of the velocities, the situation in counter-flowing superfluid component of $^4$He becomes similar to that in $^3$He turbulence with a normal fluid component at rest. Thus one expects that the spectrum~\eqref{LNV} describes the energy distribution between scales for $k\gg  k_\times$.

For $k < k_\times$ due to the partial  velocity correlations the energy dissipation is much weaker than  for $k> k_\times$,  although  it cannot be neglected as in co-flowing $^4$He, with classical Kolmogorov-1941 (K41) energy spectrum. Thus we can expect only moderate suppression of the energy spectrum as compared to the K41 case, as was recently observed in \Ref{Vin}.

A more detailed analysis of the energy spectra $\C E\sb s(k)$ and  $\C E\sb n(k)$ in the counter-flowing $^4$He that accounts for the decoupling of the normal and superfluid turbulent velocity fluctuations and the resulting energy dissipation due to the mutual friction is beyond the scope of this paper.
%============================================================
\acknowledgements  We acknowledge L. Skrbek, S. Babuin and  E. Varga for numerous and useful discussions of similarities and differences  between co- and counter-flowing turbulence of superfluid $^4$He that inspired current research.
VSL and AP acknowledge kind hospitality in Prague university and support of EuHIT project ``V-Front" that make their visit possible.
%==============================================================
\appendix
\section{\label{ss:background} Some Definitions and Known Relationships}

To find the cross-correlation  $\< \B u'\sb n \cdot \B u' \sb s\>$ we need to recall some definitions and relationships required for our derivation, which are well-known in statistical physics. The
first is the set of Fourier transforms  in the following normalization:
\begin{subequations}\label{FT}
 \begin{eqnarray}\label{FTa}%%
    \B u'\sb{n,s}(\B r,t)& \= &   \int   \frac{ d \B k}{(2\pi)^3}   \,   \B
v\sb{n,s}(\B k,t) \exp(i \B k\* \B r )\,,\\ \label{FTb}
  \B v\sb{n,s}(\B k,t) & \= &    \int    \frac{d\omega}{2\pi} \,
\~{\B v}\sb{n,s}(\B k,\omega) \exp(-i \omega t)\,,\\ \label{FTc}
  \~{\B v}\sb{n,s}(\B k,\omega)  &=&    \int    d \B r   d t\  \B
u'\sb{n,s}(\B r,t) \exp[i(\omega t-\B k\* \B r)]\ .~~~~~~~
\end{eqnarray} %%
The same normalization will be used for other objects of interest.
\end{subequations}

Next we define the simultaneous correlations and cross-correlations in $\B k$-representation, (proportional to $\d(\B k - \B k')$ due to homogeneity):
\begin{subequations}\label{corr}
\begin{eqnarray}\label{corr-nn}\< \B v\sb n(\B k,t)\*\B v^* \sb n(\B k',t)  \>&=& (2\pi)^3 E\sb{nn}(\B k)\, \d(\B k-\B k')\,,~~~ \\
\label{corr-ss}
\< \B v\sb s(\B k,t)\*\B v^* \sb s(\B k',t)  \>&=& (2\pi)^3 E\sb{ss}(\B k)\, \d(\B k-\B k')\,, \\ \label{Corr-ns}
\< \B v\sb n(\B k,t)\*\B v^* \sb s(\B k',t)  \>&=& (2\pi)^3 E\sb{ns}(\B k)\, \d(\B k-\B k')\ .
\end{eqnarray}\end{subequations}
We also need to define cross-correlations $\< \~{\B v}\sb n \*\~{\B v}^* \sb s  \> $ in $(\B k,\omega)$-representation:
\begin{subequations}\label{cross} \begin{eqnarray}\label{crossA}
&&\< \~{\B v}\sb n(\B k,\omega)\*\~{\B v}^* \sb s(\B k',\omega')  \>\\ \nn &=& (2\pi)^4 \~ E\sb{ns}(\B k,\omega)\, \d(\B k-\B k')\, \delta(\omega-\omega') \ .
\end{eqnarray}
This object is related to the simultaneous $\< \B v\sb n \*\B v^* \sb s  \> $ cross-correlation\,\eqref{Corr-ns} via the frequency integral:
\begin{equation}\label{crossA}
\< \B v\sb n(\B k,t)\*\B v^* \sb s(\B k',t)  \>=\int d\omega \~ E\sb{ns}(\B k,\omega)\ .
\end{equation}
\end{subequations}
 Here and below ``tilde'' marks the objects defined in $(\B k,\omega)$-representation.

It is known also that the $\B k$-integration of the correlations\,\eqref{corr} produces  their one-point second moment:
\begin{subequations}\label{def6}
\begin{eqnarray}\label{def6a}
 \int \frac{d \B k }{(2\pi)^3}E\sb {nn}(\B k)&=& \< |\B u\sb {n} (\B r, t)|^2 \>\,,\\ \label{def6b}
  \int \frac{d \B k }{(2\pi)^3}E\sb {ss}(\B k)&=& \< |\B u\sb {s} (\B r, t)|^2 \>\,,\\ \label{def6c}
   \int \frac{d \B k }{(2\pi)^3}E\sb {ns}(\B k)&=& \< \B u\sb {n} (\B r, t)\*
   \B u\sb s (\B r, t) \>\ .
\end{eqnarray}\end{subequations}
In the isotropic  case, each of the three correlations $E_{\dots} (\B k)$ is independent of the direction of $\B k$:  $E_{\dots} (\B
k)=E_{\dots}  (k)$ and $\int \dots d\B k= 4\pi \int \dots k^2 \, dk$.  This allows the introduction of the  one-dimensional energy spectra $\C E\sb s$, $\C E \sb n$ and the cross-correlation $\C E\sb{ns}$ as follows:
 \begin{eqnarray}\nn % \begin{eqnarray}\label{def1a}
\C E\sb{n}(k)&=& \frac{k^2}{2\pi^2}E\sb{nn}(k)\,, \quad \C E\sb{s}(k)= \frac{k^2}{2\pi^2}E\sb{ss}(k)\,,\\ \label{def1a}
\C E\sb{ns}(k)&\=& \frac{k^2}{2\pi^2}E\sb{ns}(k)\ .
\end{eqnarray}

%================================================


\begin{thebibliography}{99}

\bibitem{1}
R. J. Donnelly, Quantized Vortices in Hellium II (Cambridge
3 University Press, Cambridge, 1991)

\bibitem{2} \emph{Quantized Vortex Dynamics and Superfluid Turbulence}, edited by C.F. Barenghi, R.J. Donnelly and W.F. Vinen, Lecture Notes in Physics
    \textbf{571} (Springer-Verlag, Berlin, 2001)

\bibitem{SS-2012} L. Skrbek, K.R. Sreenivasan
 %Developed quantum turbulence and its decay.
 Phys Fluids \textbf{24} 011301 (2012).

\bibitem{BLR} C. F. Barenghi, V. S. L'vov, and P.-E. Roche,
%Experimental, numerical, and analytical velocity spectra in turbulent quantum fluid,
Proc Natl Acad Sci USA \textbf{111}, 4683  (2014).

 \bibitem{grid} M. R. Smith, R. J. Donnelly, N. Goldenfeld, and W. F.
Vinen, Phys. Rev. Lett. \textbf{71}, 2583 (1993).

\bibitem{channel1} P.L. Walstrom, J. G. Weisend, J.R. Maddocks and S. W. Van Sciver, Criogenics \textbf{28}, 101 (1988).

\bibitem{channel2} S. Babuin · E. Varga · L. Skrbek, J Low Temp Phys \textbf{175} 324  (2014).

\bibitem{Vinen} H. E. Hall and W. F. Vinen, Proc. Roy. Soc. A \textbf{238}, 204 (1956).

\bibitem{SR-1991} K.W. Schwarz and J. R. Rozen, Phys. Rev. Lett., \textbf{66}, 1896 (1991).

\bibitem{L199} V.S. L'vov, S.V. Nazarenko and L. Skrbek, J. Low Temperature Physics, \textbf{145}, 125 (2006).

\bibitem{59} P-E Roche, C.F. Barenghi, E. Leveque
%Quantum turbulence at finite temperature: The two-fluids cascade.
Europhys Lett \textbf{87} 54006 (2009).

\bibitem{60} Tchoufag J, Sagaut P
%Eddy damped quasi normal Markovian simulations of superfluid turbulence in helium II.
Phys Fluids \textbf{22} 125103 (2010).

 \bibitem{PRB2015} L. Boue, V S. L'vov., Y. Nagar., S. V. Nazarenko, A. Pomyalov., and I. Procaccia.
    %Energy and Vorticity Spectra in Turbulent Superfluid 4He from T = 0 to T-lambda.
    Phys. Rev. B \textbf{91}, 144501 (2015).


\bibitem{61} L. Boue, V.S. L'vov, A. Pomyalov, I. Procaccia  
%Enhancement of intermittency in superfluid turbulence.
 Phys Rev Lett \textbf{110} 014502 (2013).
\bibitem{24} W. F. Vinen, J. Low Temp. Phys., \textbf{175}  305 (2014).

\bibitem{BK61}I.L. Bekarevich, I.M. Khalatnikov,%Phenomenological derivation of the equations of vortex motion in He II,
 Sov. Phys. JETP \textbf{13} (3), 643-646 (1961).


    %%%%review

 \bibitem{VinenNiemela} W.~F. Vinen and J.~J. Niemela,   J. Low Temp. Phys. {\bf 128}, 167 (2002)

\bibitem{LNV} V.~S. L'vov, S.~V. Nazarenko, G.~E. Volovik, JETP Letters \textbf{80}, 479 (2004). % # 16
\bibitem{19} K. W. Schwarz, Phys. Rev. B \textbf{38}, 2398 (1988).

\bibitem{BLPP-2013} L. Kondaurova, V. S. L'vov, A. Pomyalov and I. Procaccia,%Kelvin waves
 Phys. Rev. B \textbf{90} 094501 (2014)

\bibitem{Kra} R. H. Kraichnan, J. Fluid Mech. \textbf{5},497 (1959).

\bibitem{Wyld} H. W. Wyld, Ann. Phys. (N. Y.) \textbf{14}, 143 (1961).
\bibitem{acoustic} V.S. L'vov, Yu. L'vov, A.C. Newell and V.E. Zakharov. %Statistical description of Acoustic Turbulence ,
Phys. Rev. E. \textbf{56}, 390 (1997).

\bibitem{LP} V.S. L'vov and I. Procaccia, Phys. Rev. E, \textbf{52}, 3840 (1995) and \textbf{52}, 3858 (1995).


\bibitem{Kond} L. Kondaurova, V. S. L'vov, A. Pomyalov and I. Procaccia, % Structure of quantum vortex tangle in He-4 counterflow turbulence",
Phys. Rev. B, \textbf{89}, 014502 (2014).



\bibitem{245} L. Boue, V S. L'vov., Y. Nagar., S. V. Nazarenko, A. Pomyalov., and I. Procaccia.
%Energy and Vorticity Spectra in Turbulent Superfluid 4He from T = 0 to T-lambda.
Phys. Rev. B \textbf{91}, 144501 (2015).


%\bibitem{Prague} \green{S. Babuin, V.S. L'vov, A. Pomyalov, L. Skrbek, E. Varga, \emph{Coexistence and interplay of quantum and classical turbulence of superfluid He-4} arXiv:1509.03765.}


\bibitem{DonnelyBarenghi98} R. J. Donnelly and C. F. Barenghi,J. Phys. Chem. Ref. Data, {\bf 27}, No. 6, 1217(1998)


 \bibitem{Sabra1} L. Boue, V.S. L'vov, A. Pomyalov, I. Procaccia,
%Energy spectra of superfluid turbulence in He-3B,
Phys. Rev. B \textbf{85}, 104502 (2012)



 \bibitem{Vin} A. Marakov, J. Gao, W. Guo, S. W. Van Sciver,G. G. Ihas, D. N. McKinsey, and W. F. Vinen, Phys. Rev B \textbf{91} 094503 (2015).












\end{thebibliography}
\end{document}